\documentclass[conference]{IEEEtran}

\usepackage{epsfig,rotating,setspace,latexsym,amsmath,epsf,amssymb,amsfonts, bm,theorem}
\usepackage{cite,color,subfigure}

\newtheorem{lemma}{Lemma}

\newenvironment{Proof}[1]{\medskip\par\noindent{\bf Proof:\,}\,#1}{{\mbox{\,$\blacksquare$}\par}}

\begin{document}

\IEEEoverridecommandlockouts

\title{Optimal Scheduling for Energy Harvesting Transmitters with Hybrid Energy Storage\thanks{This work was supported by NSF Grants CNS 09-64632, CCF 09-64645, CCF 10-18185 and CNS 11-47811.}}

\author{\IEEEauthorblockN{Omur Ozel \qquad Khurram Shahzad \qquad  Sennur Ulukus}
\IEEEauthorblockA{Department of Electrical and Computer Engineering\\
University of Maryland College Park, MD 20742\\
{\it omur@umd.edu \qquad kshahzad@umd.edu \qquad  ulukus@umd.edu}}}

\maketitle

\begin{abstract}
We consider data transmission with an energy harvesting transmitter which has a hybrid energy storage unit composed of a perfectly efficient super-capacitor (SC) and an inefficient battery. The SC has finite space for energy storage while the battery has unlimited space. The transmitter can choose to store the harvested energy in the SC or in the battery. The energy is drained from the SC and the battery simultaneously. In this setting, we consider the offline throughput maximization problem by a deadline over a point-to-point channel. In contrast to previous works, the hybrid energy storage model with finite and unlimited storage capacities imposes a generalized set of constraints on the transmission policy. As such, we show that the solution generalizes that for a single battery and is obtained by applying directional water-filling algorithm multiple times. 
\end{abstract}

\section{Introduction}

A key determinant of the performance of energy management policies in energy harvesting systems is the efficiency of energy storage. Energy storage units may foster imperfections and a well-known design method to boost the energy storage efficiency is to augment a super-capacitor (SC) to the existing battery and obtain a hybrid energy storage unit, see e.g., \cite{solar,solar2}. In this literature, it is common knowledge that super-capacitors can store energy nearly ideally; however, they suffer from low energy storage capacities. On the other hand, batteries have large storage capacities while they suffer from inefficient energy storage. In this paper, we consider scheduling the data transmission in energy harvesting transmitters with such a hybrid energy storage unit. 

In data transmission with such a device, aside from determining the transmit power level, the transmitter has to decide the portions of the incoming energy to be saved in the SC and the battery. While it is desirable to save energy in the SC due to its perfect storage efficiency, the storage capacity limitation necessitates careful management of the energy saved in this device. In this regard, the transmitter may wish to save energy in the inefficient battery rather than losing it. Therefore, the extra degree of freedom to choose the portions of incoming energy to save in different storage units significantly complicates the energy management problem. In this paper, we address this problem in an offline setting.    

Offline throughput maximization for energy harvesting systems has recently received considerable interest \cite{tcom-submit, kaya_subm, ozel11, wless-submit, uysal_paper, finite, jingmac, kaya_jcn, Zhang_Relay, gunduz_camsap, Gunduz-leakage,Gurakan12ISIT,asilomar,icc2013,jiexu12,Orhan12ITW, kaya_subm2}. In \cite{tcom-submit}, the transmission completion time minimization problem is solved in energy harvesting systems with an unlimited capacity battery that operates over a static channel. The solution of this problem has later been extended for a finite capacity battery \cite{kaya_subm}, fading channel \cite{ozel11}, broadcast channel \cite{wless-submit,uysal_paper,finite}, multiple access channel \cite{jingmac}, interference channel \cite{kaya_jcn} and relay channel \cite{Zhang_Relay,gunduz_camsap}. Offline throughput maximization for energy harvesting systems with leakage in energy storage was studied in \cite{Gunduz-leakage}. In \cite{Gurakan12ISIT,asilomar,icc2013}, offline optimal performance limits of multi-user wireless systems with energy transfer are studied. This literature has also been extended in \cite{jiexu12,Orhan12ITW} for systems with processing costs, which is another common non-ideal behavior for these systems. Finally, \cite{kaya_subm2} considers offline throughput maximization for energy harvesting devices in the presence of energy storage losses.   

Previous works on offline throughput maximization did not address the hybrid energy storage model; however, a two-unit storage model in this spirit has appeared in \cite{Zhang12ISIT}. In this reference, the authors analyze a save-then-transmit protocol in energy harvesting wireless systems with main and secondary energy storage devices that operate over fading channels. The objective is to minimize the outage probability over a single variable, namely the save ratio. Using this analysis, some useful guidelines are given. Our work is different from \cite{Zhang12ISIT} in that our objective is throughput maximization and we perform the optimization over a sequence of variables. Moreover, unlike our hybrid storage model, both of the storage devices have unlimited capacities in the model of \cite{Zhang12ISIT}.

In this paper, we address the offline throughput maximization problem for the specified hybrid energy storage model. As emphasized in \cite{tcom-submit, kaya_subm, ozel11, wless-submit, uysal_paper, finite, jingmac, kaya_jcn, Zhang_Relay, gunduz_camsap, Gunduz-leakage,Gurakan12ISIT,asilomar,icc2013, jiexu12,Orhan12ITW, kaya_subm2}, energy arrivals impose causality constraints on the energy management policy. In addition, battery limitation imposes no-energy-overflow constraints \cite{kaya_subm, ozel11, finite}. In the presence of hybrid energy storage, the energy causality and no-energy-overflow constraints take a new form. We capture the inefficiency of the battery by a factor $\eta$ and solve the resulting offline throughput maximization problem. 

A natural way of formulating this problem for the specified model is over the powers drained from the SC and the battery and the portion of the incoming energy to be saved in the SC. Instead, in the spirit of \cite{solar2}, we formulate the problem in terms of energies drained from the SC and the battery and energy transferred from the SC to the battery. This formulation reveals many commonalities of this problem with the previous works. This problem relates to sum-throughput maximization in a multiple access channel with energy harvesting transmitters \cite{jingmac} since energies drained from two queues contribute to transmission of a common data. Battery storage loss model is reminiscent of that in \cite{kaya_subm2} where the transmitter is allowed to save the incoming energy in a lossy battery or use it immediately for data transmission. Finally, one-way energy transfer from the SC to the battery relates to the problem considered in \cite{asilomar} where a two-user multiple access channel is considered with energy transfer from one node to the other. 

Despite the coupling between the variables that represent energies drained from and transferred within the energy storage unit, we show that the problem can be solved by application of directional water-filling algorithm \cite{ozel11} in multiple stages. In particular, we first forbid energy transfer from the SC to the battery and solve this restricted optimization problem. We show that this problem is solved by optimizing the SC allocation first and then the battery allocation given the SC allocation. Next, we allow energy transfer from the SC to the battery and show that the optimal allocation is obtained by directional water-filling in a setting transformed by the storage efficiency $\eta$. As a consequence, we obtain a generalization of the directional water-filling algorithm which yields useful insight on the structure of the optimal offline energy allocation in energy harvesting systems. 

\section{System Model}

\begin{figure}[t]
\begin{center}
\includegraphics[width=0.82\linewidth]{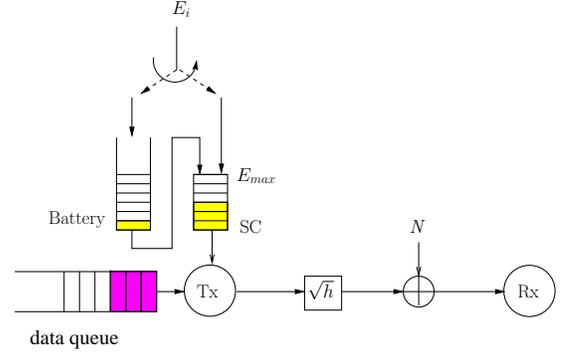}
\end{center}\vspace{-0.1in}
\caption{System model with hybrid energy storage.}\vspace{-0.2in}
\label{sys1}
\end{figure}

We consider a single-user additive Gaussian noise channel with an energy harvesting transmitter. The transmitter has three queues: a data queue and two energy queues. Two energy queues correspond to a hybrid energy storage unit composed of a battery and a super-capacitor (SC) as shown in Fig.~\ref{sys1}. The battery has unlimited storage capacity whereas SC can store at most $E_{max}$ units of energy. The battery is inefficient in the sense that the energy that can be drained from it is less than the amount that is stored; the SC is perfectly efficient. We assume infinite backlog in the data queue.  

The physical layer is an AWGN channel with the input-output relation $Y=\sqrt{h}X + N$ where $h$ is the squared channel gain and $N$ is Gaussian noise with zero-mean and unit-variance. Without loss of generality, we set $h=1$ throughout the communication. We follow a continuous time model and instantaneous rate is
\begin{align}
\label{rp}
r(t)=\frac{1}{2}\log\left(1+p(t)\right)
\end{align} 

At time $t_i^e$, $E_i$ amount of energy arrives. $E_0^b$ and $E_0^{sc}$ amounts of energies are available at the beginning in the battery and in the SC, respectively. In the following, we refer to the time interval between two energy arrivals as an epoch. More specifically, epoch $i$ is the time interval $[t_i^e,t_{i+1}^e)$ and the length of the epoch $i$ is $\ell_i=t_{i+1}^e-t_i^e$.

Whenever energy $E_i$ arrives at time $t_i^e$, the transmitter stores $E_i^b$ amount in the battery and $E_i^{sc}=E_i-E_i^b$ amount in the SC. Since SC can store at most $E_{max}$ units of energy, $E_i^{sc}$ must be chosen such that no energy unnecessarily overflows. For this reason, $E_i^{sc} \leq E_{max}$ must necessarily be satisfied. The efficiency of the battery is given by the parameter $\eta$ where $0 \leq \eta < 1$: If $E_i^b$ units of energy is stored in the battery, then $\eta E_i^b$ units can be drained and $(1-\eta)E_i^b$ units are lost. Moreover, we assume that the available energy in the battery can be transferred to SC instantaneously\footnote{In real systems, switching time between the battery and the SC is very small compared to epoch lengths of interest \cite{solar}.}. As a consequence, none of the arrived energy overflows; however, there is an energy loss due to inefficiency of the battery.  

A transmit power policy is denoted as $p(t)$ over $[0,T]$. $p(t)$ is constrained by the energy that can be drained from the hybrid storage system:
\begin{align}
\label{c1}
\int_{0}^{t_i^e} p(u)du \leq \sum_{j=0}^{i-1} E_j^{sc} + \eta E_j^{b}, \qquad \forall i 
\end{align}
where $t_i^e$ in the upper limit of the integral is considered as $t_i^e - \epsilon$ for sufficiently small $\epsilon$. 

Moreover, we note that the power policy should cause no energy overflow in the SC. In order to express this constraint, we divide each incremental drained energy $p(u)du$ as a linear combination of the energy drained from the SC, $p^{sc}(u)du$, and the energy drained from the battery, $p^b(u)du$. That is, $p(u)du = p^{sc}(u)du + p^b(u)du$. We are allowed to divide $p(u)du$ into such components since the energy in the battery can be instantaneously transferred to the SC. No-energy-overflow constraint in the SC can now be expressed as follows:
\begin{align}
\label{c2}
\sum_{j=0}^{i} E_j^{sc} - \int_{0}^{t_i^e} p^{sc}(u)du \leq E_{max}
\end{align}
We note that the constraints in (\ref{c1}) and (\ref{c2}) generalize the energy causality and no-energy-overflow constraints in the single-stage energy storage models studied, e.g., in \cite{ozel11}.

\section{Offline Throughput Maximization Problem}

In this section, we consider the offline throughput maximization problem by a deadline $T$. We first note that the power policy $p(t)$ has to take a constant value over each epoch, due to the concavity of the rate-power relation in (\ref{rp}). Therefore, the power policy is represented by the sequence of power levels $p_i=p_i^{sc}+p_i^b$ where $p_i^{sc}$ and $p_i^b$ are the portions of the power drained from the SC and the battery, respectively, in epoch $i$. Moreover, the transmitter decides the portions of the incoming energy $E_i^{sc}$ and $E_i^{b}$ so that $E_i^{sc}+E_i^b=E_i$. Since the battery is inefficient ($0 \leq \eta < 1$), we prefer to initially allocate incoming energy to the SC and the remaining energy to the battery while still allowing to transfer a portion of the initially stored energy in SC to the battery. We denote the energy transfer power at epoch $i$ as $\delta_i$ with the convention that the transferred energy becomes available for use in epoch $i+1$. In view of (\ref{c1})-(\ref{c2}), we get the following constraints for all $i$:
\begin{align} 
\label{xx1}
\sum_{j=1}^{i} \left(p_{j}^{sc} \ell_j+ \delta_{j}\ell_{j}\right) &\leq  \sum_{j=0}^{i-1} E_j^{sc}\\
\sum_{j=0}^{i} E_{j}^{sc} - \sum_{j=1}^{i} \left(p_{j}^{sc}\ell_j+\delta_{j}\ell_{j}\right)  &\leq E_{max}\\
\sum_{j=1}^{i} p_{j}^b \ell_j &\leq  \sum_{j=0}^{i-1} \left( \eta E_j^b + \eta \delta_{j}\ell_{j} \right)\\
p_i^{sc} \geq 0, \ p_i^{b} \geq 0, \ \delta_i &\geq 0 \label{xx4} 
\end{align}
where $E_i^{sc}=\min\{E_i, E_{max}\}$ and $E_i^{b}=\left(E_i - E_{max} \right)^+$. We set $\delta_0=0$ and $\delta_N=0$ by convention. We remark that in the system model, energy transfer from SC to the battery is not allowed. However, due to the offline nature, we have the freedom to allocate energy to SC first and then transfer it to the battery. Moreover, one epoch delay in the energy transfer emphasizes the fact that if the energy in the SC in epoch $i$ is transferred to the battery, that energy must be utilized starting from epoch $i+1$ as otherwise such an energy transfer cannot increase the throughput since the battery is inefficient. 

Offline throughput maximization problem by deadline $T$ with the hybrid energy storage unit is:
\begin{eqnarray}
\nonumber
\max_{p_{i}^{sc},p_{i}^{b},\delta_i\geq 0} & & \sum_{i=1}^{N} \frac{\ell_i}{2}\log\left(1 + p_{i}^{sc}+p_{i}^b\right)\\
\mbox{s.t.} & & (\ref{xx1})-(\ref{xx4}) \label{opt1}
\end{eqnarray}
We note that the problem in (\ref{opt1}) is a convex optimization problem and we can solve it using standard techniques \cite{boyd}. The Lagrangian function for (\ref{opt1}) is
\begin{align}\nonumber
\mathcal{L}=&-\sum_{i=1}^N \frac{\ell_i}{2}\log\left(1+p_{i}^{sc}+p_{i}^{b}\right) \\ \nonumber &+ \sum_{i=1}^N \lambda_i\left(\sum_{j=1}^i \left(p_{j}^{sc} \ell_j+ \delta_{j}\ell_{j}\right) -  \sum_{j=0}^{i-1} E_{j}^{sc} \right) \\ \nonumber &+\sum_{i=1}^{N-1}\mu_i\left(\sum_{j=0}^{i} E_{j}^{sc} - \sum_{j=1}^{i} \left(p_{j}^{sc}\ell_j+ \delta_{j}\ell_{j}\right) - E_{max}\right) \\ &+ \sum_{i=1}^N \nu_i\left(\sum_{j=1}^{i} p_{j}^b \ell_j -  \sum_{j=0}^{i-1} \left(\eta E_{j}^b + \eta \delta_{j}\ell_{j}\right)\right) \nonumber \\  &- \sum_{i=0}^N \gamma_i \delta_{i} - \sum_{i=1}^N \rho_{1i} p_{i}^{sc} - \sum_{i=1}^N \rho_{2i} p_{i}^b \label{lagrangian} 
\end{align}

KKT optimality conditions for (\ref{opt1}) are:
\begin{align}\label{kkt1}
-\frac{1}{1+p_{i}^{sc}+p_{i}^b} + \sum_{j=i}^N \lambda_j - \sum_{j=i}^{N-1} \mu_j - \rho_{1i} &= 0, \quad \forall i \\
-\frac{1}{1+p_{i}^{sc}+p_{i}^b} + \sum_{j=i}^N \nu_j - \rho_{2i} &= 0, \quad \forall i \label{kkt2} \\ \sum_{j=i}^N \lambda_j - \sum_{j=i}^{N-1} \mu_j - \eta\sum_{j=i+1}^N \nu_j - \gamma_i &= 0, \quad \forall i \label{kkt3}
\end{align}
and the complementary slackness conditions are:
\begin{align}
\label{slack1}
\lambda_i\left(\sum_{j=1}^i \left(p_{j}^{sc} \ell_j+ \delta_{j}\ell_{j}\right) -  \sum_{j=0}^{i-1} E_{j}^{sc} \right) &= 0, \ \ \forall i \\
\mu_i\left(\sum_{j=0}^{i} E_{j}^{sc} - \sum_{j=1}^{i} \left(p_{j}^{sc}\ell_j+ \delta_{j}\ell_{j}\right) - E_{max}\right)&=0, \ \ \forall i \\ \nu_i\left(\sum_{j=1}^{i} p_{j}^b \ell_j -  \sum_{j=0}^{i-1}\left( \eta E_{j}^b + \eta \delta_{j}\ell_{j}\right)\right) &= 0, \ \ \forall i \\  \gamma_i \delta_{i} = \rho_{1i} p_{i}^{sc} = \rho_{2i} p_{i}^{b} &= 0, \ \  \forall i \label{slack4}
\end{align}
We note that the optimization problem (\ref{opt1}) may have many solutions. In order to get a solution, it suffices to find power sequences $p_{i}^{sc}$, $p_{i}^b$ and Lagrange multipliers that are consistent with (\ref{kkt1})-(\ref{kkt3}) and (\ref{slack1})-(\ref{slack4}). We observe properties of an optimal solution $p_{i}^{sc*},p_{i}^{b*}$ and $\delta_i^*$ in the following lemmas.

\begin{lemma}\label{lm1}
If $p_{i}^{b*} \neq 0$, $p_{i}^{sc*}+p_{i}^{b*}$ does not decrease in the passage from epoch $i$ to epoch $i+1$. 
\end{lemma}
\begin{Proof}
When $p_{i}^{b*} \neq 0$, we have $\rho_{2i}=0$. By (\ref{kkt2}), we have $p_{i}^{sc*}+p_{i}^{b*}=\frac{1}{\sum_{j=i}^N \nu_j}-1$ and
$p_{i+1}^{sc*}+p_{i+1}^{b*}=\frac{1}{\sum_{j=i+1}^N \nu_j - \rho_{2(i+1)}}-1$.
Since $\nu_{i}\geq 0$ and $\rho_{i+1} \geq 0$, we conclude the desired result.  
\end{Proof}

\begin{lemma} \label{lm12}
If $E_{i-1}^{b} \neq 0$, $p_{i}^{b*} = 0$ and $p_{i+1}^{b*} \neq 0$, then $p_{i}^{sc*} + p_{i}^{b*}$ does not increase in the passage from epoch $i$ to epoch $i+1$. Similarly, if $E_{i-1}^b = 0$, $E_{i}^b = 0$, $p_{i}^{b*} = 0$ and $p_{i+1}^{b*} \neq 0$, then $p_{i}^{sc*} + p_{i}^{b*}$ does not increase in the passage from epoch $i$ to epoch $i+1$.
\end{lemma}
\begin{Proof}
As $p_{i}^{b*} = 0$ and $p_{i+1}^{b*} \neq 0$, we have $\rho_{2i}\geq 0$ and $\rho_{2(i+1)}=0$. Moreover, since $p_{i}^{b*} = 0$, $\nu_i = 0$ as the constraint $\sum_{j=1}^i p_j^b \ell_j \leq \sum_{j=0}^{i-1} (\eta E_j^b + \eta \delta_j\ell_j)$ cannot be satisfied with equality when $E_{i-1}^b \neq 0$ and $p_i^{b*}=0$. Similarly, we note that if $E_{i-1}^b = 0$, $E_{i}^b = 0$, then $\sum_{j=1}^i p_j^b \ell_j \leq \sum_{j=0}^{i-1} (\eta E_j^b + \eta \delta_j\ell_j)$ cannot be satisfied with equality when $p_i^{b*}=0$ and $p_{i+1}^{b*} \neq 0$. Therefore, $\sum_{j=i}^N \nu_j - \rho_{2i} \leq \sum_{j=i+1}^N \nu_j - \rho_{2(i+1)}$, which by (\ref{kkt2}) implies the desired result. 
\end{Proof}
\begin{lemma}\label{lm2}
If $p_{i}^{sc*},p_{i}^{b*} \neq 0$, then $\delta_{i}^*=0$.
\end{lemma}
\begin{Proof}
If $p_{i}^{sc*},p_{i}^{b*} \neq 0$, from (\ref{kkt1}) and (\ref{kkt2}), we have $\sum_{j=i}^N \lambda_j - \sum_{j=i}^{N-1} \mu_j = \sum_{j=i}^N \nu_j$. Combining this with (\ref{kkt3}), we conclude that $\gamma_{i}=\nu_i + (1-\eta)\sum_{j=i+1}^N \nu_j>0$ as $\eta < 1$. In view of the slackness condition $\gamma_{i}\delta_{i}=0$, we get $\delta_i^* = 0$.
\end{Proof}

\begin{lemma}\label{22}
If $p_i^{sc*},p_{i+1}^{sc*}, p_{i+1}^{b*} \neq 0$, $p_i^{sc*} + p_i^{b*} \leq p_{i+1}^{sc*} + p_{i+1}^{b*}$, then $\delta_i^* = 0$.
\end{lemma}
\begin{Proof}
As $p_i^{sc*},p_{i+1}^{sc*},p_{i+1}^{b*} \neq 0$, $\rho_{1i} = \rho_{1(i+1)} = \rho_{2(i+1)} = 0$. Therefore, by (\ref{kkt1}) and since $p_i^{sc*} + p_i^{b*} \leq p_{i+1}^{sc*} + p_{i+1}^{b*}$, we have $\sum_{j=i}^N \lambda_j - \sum_{j=i}^{N-1} \mu_j > \sum_{j=i+1}^N \lambda_j - \sum_{j=i+1}^{N-1} \mu_j$.
Moreover, since $\rho_{2(i+1)} = 0$, we have $\sum_{j=i+1}^N \lambda_j - \sum_{j=i+1}^{N-1} \mu_j = \sum_{j=i+1}^N \nu_j$. By (\ref{kkt3}), $\gamma_i>0$ and due to the slackness condition $\gamma_{i}\delta_{i}=0$, we get $\delta_i^* = 0$.
\end{Proof}

Lemmas \ref{lm1}-\ref{22} reveal significant properties of optimal power sequences $p_{i}^{sc*}$ and $p_{i}^{b*}$. In particular, these lemmas indicate that $p_{i}^{b}$ has to be carefully determined. Note that since energy is first allocated to the SC, $p_{i}^{sc*}>0$ for all $i$. In view of these lemmas, we adopt the following strategy: Initially, we fix $\delta_i=0$ and find the optimal policy under this constraint. Note that $\delta_i=0$ is a good candidate for an optimal selection in view of Lemmas \ref{lm2}-\ref{22}. If the resulting optimal policy is compatible with the KKT conditions, then we stop. Otherwise, we carefully update $\delta_i$ so that the KKT conditions are satisfied. 

\section{Finding the Optimal Policy for $\delta_i=0$}

For fixed $\delta_i=0$, the problem becomes maximizing the throughput by the deadline subject to energy causality and finite SC $E_{max}$ constraints only:
\begin{eqnarray}
\nonumber
\max_{p_{i}^{sc},p_{i}^{b}\geq 0} & & \sum_{i=1}^{N} \frac{\ell_i}{2}\log\left(1 + p_{i}^{sc}+p_{i}^b\right)\\ \nonumber
\mbox{s.t.} & & (\ref{xx1})-(\ref{xx4}) \\ & & \delta_i=0, \ \forall i \label{opt12}
\end{eqnarray}
where $E_i^{sc}=\min\{E_i, E_{max}\}$ and $E_i^{b}=\left(E_i - E_{max} \right)^+$. We note that (\ref{opt12}) is equivalent to sum-throughput maximization in a two-user multiple access channel with finite and infinite capacity batteries. A simpler version of this problem where both users have infinite capacity battery is addressed in \cite{jingmac}. While the problem of sum-throughput maximization has a simple solution when batteries are unlimited by summing the energies of the users and performing single-user throughput maximization \cite{jingmac}, the finite battery constraint in (\ref{opt12}) prevents such a simple solution. As in the general problem in \cite{jingmac}, the solution of (\ref{opt12}) is found by iterative directional water-filling where infinitely many iterations are required in general.

Next, we show that due to the problem structure, we can find the solution of (\ref{opt12}) only in two iterations. Note that the energy arrivals of the storage units are $E_i^{sc}=\min\{E_i, E_{max}\}$ and $E_i^{b}=\left(E_i - E_{max} \right)^+$: Energy is first allocated to the SC and the remaining energy is allocated to the battery. This specific way of allocation allows us to find the solution only in two iterations. We formally state this result in the following lemma.

\begin{lemma}
\label{lm3}
For fixed $\delta_i=0$, let $\hat{p}_{i}^{sc}$ be the outcome of directional water-filling given $p_{i}^b=0$. Let $\hat{p}_{i}^b$ be the outcome of directional water-filling given $\hat{p}_{i}^{sc}$. Then, $\hat{p}_{i}^{sc}$ and $\hat{p}_{i}^b$ are jointly optimal for (\ref{opt12}).
\end{lemma}
The proof of Lemma \ref{lm3} follows from two facts: First, since the SC has finite storage, optimal allocation can be performed independently between the epochs in which energy arrival is exactly equal to the storage capacity \cite{kaya_subm}. Second, the energy arrival of the battery is non-zero only over epochs in which energy arrival of the SC is exactly equal to its storage capacity. We leave the detailed proof for a longer version of this work. We note that the claim in Lemma \ref{lm3} would not be true if $E_{i}^{sc}$ and $E_{i}^b$ were allowed to take arbitrary values. Therefore, apart from providing a crucial step towards finding the solution of (\ref{opt1}), the result stated in Lemma \ref{lm3} is an interesting case for the two-user multiple access channel with finite battery constraints where the optimal is reached only in two iterations.  

\section{Determining The Optimal $\delta_i^*$}

We note that for $\hat{p}_{i}^{sc}$ and $\hat{p}_{i}^b$, there are Lagrange multipliers $\lambda_i$, $\mu_i$, $\nu_i$, $\rho_{1i}$ and $\rho_{2i}$ that are compatible with (\ref{kkt1}) and (\ref{kkt2}). However, there may not exist $\gamma_i \geq 0$ that are compatible with (\ref{kkt3}). In this section, we propose a method to update the allocations $\hat{p}_{i}^{sc}$ and $\hat{p}_{i}^b$ and the Lagrange multipliers $\lambda_i,\mu_i,\nu_i,\rho_{1i},\rho_{2i}$ that yield $\delta_i^*$ and corresponding $\gamma_i$ so that (\ref{kkt1})-(\ref{kkt3}) and (\ref{slack1})-(\ref{slack4}) are satisfied. For brevity, we restrict our treatment to the case where $E_1^b>0$ and $E_i^b = 0$ for $i=2,\ldots,N$; however, the arguments can be easily generalized. One can show that in this case, $\nu_N>0$ and $\nu_i=0$ for $i=1,\ldots,N-1$.  

\begin{figure}[t]
\begin{center}
\includegraphics[width=0.85\linewidth]{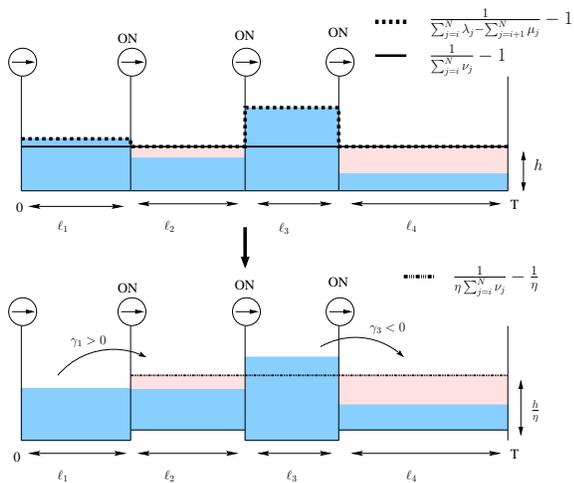}
\end{center}\vspace{-0.15in}
\caption{Transforming the directional water-filling setting.}\vspace{-0.1in}
\label{sys3}
\end{figure}

Note that if $\hat{p}_{i}^b \neq 0$ for some $i$, resulting Lagrange multipliers yield $\gamma_i \geq 0$. In view of the KKT condition (\ref{kkt3}), we transform the directional water-filling setting as in Fig. \ref{sys3}: We multiply the water level and the bottom level by $\frac{1}{\eta}$ at epochs where $\hat{p}_{i}^b > 0$ and leave other epochs unchanged where the bottom level is $1$. Moreover, if $\gamma_i\geq 0$, we set $\delta_i^*=0$ and transform the water level and the bottom level of that epoch. At epochs $i$ with $\gamma_i<0$, we wish to decrease $\sum_{j=i}^N\nu_j$ and increase $\sum_{j=i}^N \lambda_j - \sum_{j=i}^{N-1}\mu_j$ so that $\gamma_i$ approaches zero and resulting allocations are compatible with (\ref{kkt1})-(\ref{kkt3}) and (\ref{slack1})-(\ref{slack4}). We next argue that if energy is transferred from epochs $i$ with $\gamma_i<0$ in a coordinated fashion, this is possible.  

Recall that $\nu_N>0$ and $\nu_i=0$ for $i=1,\ldots,N-1$. We decrease $\nu_N$ and increase $\lambda_{\tilde{i}}$, $\mu_{\tilde{i}}$ and $\sum_{j=\tilde{i}}^N \lambda_j - \sum_{j=\tilde{i}}^{N-1}\mu_j$ where $\tilde{i}$ is the epoch index with the lowest $\sum_{j=i}^N \lambda_j - \sum_{j=i}^{N-1}\mu_j$. This decreases the power level $p^{sc}_{\tilde{i}}$ and increases the battery power level $p_i^b$ at all epochs. Therefore, a non-zero energy transfer from epoch $\tilde{i}$ occurs. As we decrease $\nu_N$, $\gamma_i$ also increases. In particular, $\gamma_i$ may change sign from negative to positive in which case, we make sure that $\delta_i^*=0$ for that epoch and hence we transform the bottom levels and the water levels for those epochs as in Fig. \ref{sys3}. On the other hand, $\sum_{j=\tilde{i}}^N \lambda_j - \sum_{j=\tilde{i}}^{N-1}\mu_j$ increases and it may hit the second lowest $\sum_{j=i}^N \lambda_j - \sum_{j=i}^{N-1}\mu_j$. In this case, we start to increase $\lambda_i$, $\mu_i$ and $\sum_{j=i}^N \lambda_j - \sum_{j=i}^{N-1}\mu_j$ in both of these epochs. 

\begin{figure}[t]
\begin{center}
\includegraphics[width=0.8\linewidth]{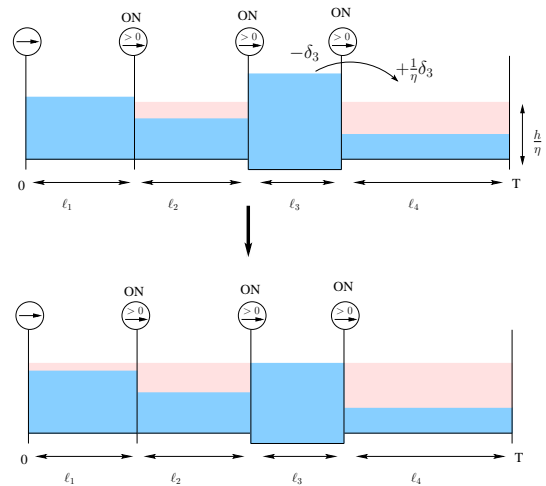}
\end{center}\vspace{-0.15in}
\caption{The water flow in the transformed directional water-filling setting.}\vspace{-0.15in}
\label{sys4}
\end{figure}

Note that this procedure corresponds to a coordinated energy transfer: We start energy transfer from the epoch $\tilde{i}$ with the highest power level $\hat{p}_i^{sc}$. In the transformed setting, as we transfer $\delta_i$, $\frac{1}{\eta} \delta_{i}$ units of water is added to the next epoch as shown in Fig. \ref{sys4}. If the power level of epoch $\tilde{i}$ decreases to the level of the second highest power $\hat{p}_i^{sc}$ with $\gamma_i < 0$, then energy is transferred simultaneously from these epochs. Causality conditions may forbid decreasing $\nu_N$ after some level. In this case, while decreasing $\nu_N$ we increase $\nu_i$ at epochs $i$ the causality condition is violated. This way, all epochs $i$ which have initially $\gamma_i<0$ are updated so that $\gamma_i \geq 0$ with $\gamma_i = 0$ if $\delta_i>0$ and (\ref{kkt1})-(\ref{kkt3}) and (\ref{slack1})-(\ref{slack4}) are satisfied.    

Note that when energy is transferred from the SC to the battery in epoch $i$, this energy spreads over future epochs $i+1,\ldots,N$. Moreover, the energy that was transferred from epochs $1,\ldots,i-1$ in the second directional water-filling of Lemma \ref{lm3} given $\hat{p}_{i}^{sc}$ may flow back to these epochs. We, therefore, measure the transferred energy within the battery at each epoch by means of meters and negate it if energy flows in the opposite direction. This is reminiscent of the meters used for the two-way channel in \cite{asilomar,icc2013}.

\section{Conclusion}

We studied offline throughput maximization in an energy harvesting transmitter with hybrid energy storage. The solution generalizes the directional water-filling algorithm in \cite{ozel11} and provides useful insights on optimal power allocation under battery limitation. As a byproduct, we obtain new insights about optimal policies over multiple access channels with and without energy transfer under finite battery constraints.

\end{document}